\documentstyle[preprint,aps]{revtex}

\begin{document}
\draft

\newcommand{\beq}{\begin{equation}}
\newcommand{\eeq}{\end{equation}}
\newcommand{\ben}{\begin{eqnarray}}
\newcommand{\een}{\end{eqnarray}}
\newcommand{\bea}{\begin{array}}
\newcommand{\eea}{\end{array}}
\newcommand{\om}{(\omega )}
\newcommand{\bef}{\begin{figure}}
\newcommand{\eef}{\end{figure}}
\newcommand{\leg}[1]{\caption{\protect\rm{\protect\footnotesize{#1}}}}

\newcommand{\ew}[1]{\langle{#1}\rangle}
\newcommand{\be}[1]{\mid\!{#1}\!\mid}
\newcommand{\no}{\nonumber}
\newcommand{\etal}{{\em et~al }}
\newcommand{\geff}{g_{\mbox{\it{\scriptsize{eff}}}}}
\newcommand{\da}[1]{{#1}^\dagger}
\newcommand{\cf}{{\it cf.\/}\ }
\newcommand{\ie}{{\it i.e.\/}\ }

\title{\center{Single photon generation by pulsed excitation of a single dipole }}

\author{Rosa Brouri, Alexios Beveratos,
Jean-Philippe Poizat, and Philippe Grangier}
\address{Laboratoire Charles Fabry de l'Institut  d'Optique, UMR 8501 du CNRS, \\
B.P. 147,
 F91403 Orsay Cedex - France\\
e-mail : jean-philippe.poizat@iota.u-psud.fr}
\maketitle
\centerline{\today}
\begin{abstract}

The fluorescence  of a single dipole excited by an intense
light pulse  can lead to the generation of another light pulse containing 
a single photon.
The influence of the duration and energy of the excitation pulse
on the number of photons in the fluorescence pulse is studied.
The case of a two-level dipole with strongly damped coherences 
is considered. 
The presence of a  metastable state leading to shelving is also 
investigated. 

\end{abstract}

\pacs{PACS. 42.50.Dv, 03.67.Dd, 33.50.-j}

\section{Introduction}

The security of quantum cryptography is based on the fact that
each bit of information is coded on a single quantum object,
namely a single photon. The fundamental impossibility of
duplicating the complete quantum state of a single particule
prevents any potential eavesdropper from intercepting the message
without the receiver noticing \cite{TRG}. In this context, the
realization of a efficient and integrable light source delivering
a periodic train of pulses containing one and only one photon,
would be an important advantage \cite{L}. 

The purpose of this paper is
to evaluate the reliability of such a source.
Assuming that a smart eavesdropper can get the information as soon as 
the number $n$ of photons in the pulse is larger than two 
(see Appendix 1), we define a fractional information leakage
 $f_{il}$ as :
\begin{equation}
\label{fil}
f_{il}=\frac{P_{n \geq 2}}{P_{n \geq 1}}
\end{equation}
where $P_{n \geq 1}$,  $P_{n \geq 2}$ are  respectively the
probabilities to get at least one and at least two photons. 
The value of $f_{il}$ has to be close to zero, while the
probability $P_e = P_{n \geq 1}$ to emit one photon during the sampling period
should be as high as possible; we note that
the probability to get exactly one photon is $(1-f_{il}) P_e$. 
For a poissonian light source, we have:
\begin{equation}
\label{poisson}
f_{il} = 1 - (1-P_e^{-1})\ln(1-P_e)
\bea{ccc}\makebox[0pt]{} \\ \simeq  \\ \mbox{}^{P_e \ll 1} \eea
\frac{P_e}{2}
\label{filPoisE}
\end{equation}
It is therefore possible to have a good reliability with
an attenuated poissonian light source, but $P_e$ has to be very small,
which makes the source quite inefficient.
A better way to have both a good reliability and high
emission probability is to design a device with fully controlled quantum
properties, able to emit truly single 
photon \cite{MGM,MJM,LK,KJRT,BLTO,KBKY,KHBR,FSZHW,BBPG}. 
One possibility to perform such an emission is to use the
fluorescence of a single dipole (e.g. a single molecule or a single
colored center). 
As a single dipole cannot emit more than one
photon at a time leading to antibunching in the photon 
statistics of the fluorescence light \cite{KDM,DW,BMOT}
 a pulsed excitation
of the dipole can be expected to produce individual photons on demand \cite{MGM}. 

In previous works \cite{MJM,LK} the emitting dipole 
was generally considered to be a radiatively damped two-level system.
In the present paper we will rather consider emitting dipoles with strongly
damped coherences, as it is the case for single molecules \cite{FSZHW} or
 single colored center \cite{BBPG} at room temperature. 
The decay time of coherences,  associated to non
radiative processes which occur in the picosecond range, 
is thus much shorter than the population decay time, 
that is typically in the 10 ns range. 
On the other hand, systems such as molecules or 
 coloured centers often have an extra metastable 
state, which is very long lived and thus can 
induce ``shelving" in the emission process. 
In order to describe these features, we will model
the emitting dipole using the three level scheme 
shown in Fig. \ref{levels}. 

Owing to the fast damping of coherences,
only level populations $\sigma_{aa}$ will be considered,
and the system's dynamics will be described
by using rate equations between the three levels. 
The system can be excited from ground
state $|1\rangle$ to excited state $|2\rangle$ 
with a pumping rate $r$. The decay rate from level $|2\rangle$ 
to level $|1\rangle$ is
$\Gamma$, but the system can also decay to a metastable state $|3\rangle$
at rate $\beta\Gamma$. The branching ratio
$\beta/(1 + \beta)$ is usually (but not necessarily) very small.
The emission rate from the metastable state will be neglected
(\ie no photons are emitted from level $|3\rangle$), but
we will assume that the system can go back from
level $|3\rangle$ to level $|2\rangle$ with a rate $r_d$.
This ``deshelving" effect has been observed experimentally \cite{DFTJKNW}, and may be important
under strong pumping conditions.  

The purpose of the present calculation is to evaluate
the efficiency of such a system in converting 
a train of classical light pulses into a train of single 
photon pulses (``photon gun") \cite{KHBR} . We will thus assume that this system
is excited by a train of light pulses of duration $\delta T$,
such that $\Gamma \delta T \ll 1$. The separation between the pulses 
is denoted by  $T$, with $\Gamma T \gg 1$. 
For ideal efficiency of the source, the dipole should be coupled
to a field mode in a microcavity, which is then damped to the outside world.
Here we will only consider free-space emission of the dipole,
assuming that the emitted light is collected
by purely passive ways, such as a parabolic retroreflector \cite{BLTO}.
The corresponding imperfect detection efficiency will
be included in the present model, but the possible effect of a microcavity
will not be considered here.

In the following section, we will introduce  a useful framework
for carrying out the calculation. Then we will
evaluate  the quantities of interest, taking into
account the detection efficiency. Finally we will present 
numerical results illustrating the behaviour of the system.

\section{Theoretical model}

\subsection{General framework}

The evolution of the populations will be described using 
the diagonal terms  of the density matrix
$\sigma_{bb}(t,t_0;a)$, which denotes the population of level $b$
at time $t$, starting from
level $a$ at time $t_0$ (where a and b may take any value from 1 to 3). 
For the following it will be convenient to define
the probability $\sigma^{(n)}_{bb}(t,t_0;a)$ to go from
state $|a\rangle$ at time $t_0$ to state $|b\rangle$ at time t,
with the emission of exactly $n$ photons.  
The quantities $\sigma^{(n)}_{bb}$ are linked to
the populations $\sigma_{bb}$ by the relation :
\begin{equation}\label{density_matrix}
  \sigma_{bb}(t,t_{0};a)=\sum_{n=0}^{\infty}\sigma^{(n)}_{bb}(t,t_{0};a)
\end{equation}
The probability density to emit one and only one photon at time t
when the system is in the state $|a\rangle$ at time $t_0$ is
given by the probability $\sigma^{(0)}_{22}(t,t_0;a)$ to be
in the excited state at time t without any photon emission :
\begin{equation}\label{one_photon}
  p_1=\Gamma \sigma_{22}^{(0)}(t,t_0;a)
\end{equation}
The quantities $\sigma^{(n)}_{bb}(t,t_0;a)$ introduced previously can
be related through the following recurrence relationship:
\begin{equation}\label{n_photon}
\sigma^{(n+1)}_{bb}(t,t_{0};a)=\int^{t}_{t_0}\Gamma
\sigma_{22}^{(n)}(t^{'},t_0;a)\sigma_{bb}^{(0)}(t,t^{'};1)
dt'
\end{equation}
In other terms, in order to emit $(n+1)$ photons, the system has to emit
the photon $(n+1)$ at time $t'$, and to emit no photon from $t'$
to $t$. The rate equations for $\sigma^{(0)}_{bb}$ can be written :
\begin{equation}\label{rate1}
\frac{\partial\sigma_{bb}^{(0)}}{\partial t}(t,t_0;a) =
\sum_{c}r_{cb}^{(0)}\sigma_{cc}^{(0)}(t,t_0;a)
\end{equation}
where similar equations hold for $\sigma_{bb}$, with coefficients $r_{cb}$.
Using equations
\ref{density_matrix}, \ref{n_photon} and \ref{rate1}, 
it can be shown (see Appendix 2) that the rate coefficients $r_{cb}^{(0)}$ 
are related  to the coefficients $r_{cb}$ by:
\begin{equation}\label{coef}
  r^{(0)}_{cb}=r_{cb}-\Gamma \delta_{b1} \delta_{c2}
\end{equation}
where $\delta_{ai}$ is one if $a=i$, and zero otherwise.
For the three-level system we are considering in figure
\ref{levels}, we thus obtain the following rate equations:
\ben
\label{equation-evolution}
\dot{\sigma}_{22}^{(0)} &=&
 r \sigma_{11}^{(0)}-(1+\beta)\Gamma \sigma_{22}^{(0)} +r_{d} \sigma_{33}^{(0)} \\
\label{equation-evolution1}
    \dot{\sigma}_{33}^{(0)} &=&
 -\Gamma_{T} \sigma_{33}^{(0)}+\beta\Gamma \sigma_{22}^{(0)}-r_{d} \sigma_{33}^{(0)}  \\
\label{equation-evolution2}
    \dot{\sigma}_{11}^{(0)} &=& -r \sigma_{11}^{(0)}+\Gamma_{T}\sigma_{33}^{(0)} 
\een
The difference between eq. \ref{equation-evolution} and the
original rate equations for populations is the missing
term proportional to $\sigma_{22}^{(0)}$ in the last equation. 
This means that the ground level
is no more filled after the emission of one photon, and ensures
the uniqueness of the emitted photon.

 Equations \ref{equation-evolution}-\ref{equation-evolution2} allow, with the
knowledge of the initial state, to derive the different
quantities of interest. We first consider in subsection
\ref{sec_a} the ideal situation of perfect collection efficiency. Subsection
\ref{sec_b} deals with non-unity  collection efficiency, which
substantially modifies results of subsection \ref{sec_a}. Finally we take
into account the effect of the metastable state in subsection
\ref{sec_c}.

\subsection{Two-level approximation with unit quantum efficiency}
\label{sec_a}

We assume first that all the emitted photons
are detected (unit quantum efficiency),
and that the dipole is initially in its ground state $| 1 \rangle$.
Assuming also that $\beta \ll 1$, we can neglect
the probability for the system to go to the metastable state in
the time interval between two excitation light pulses, and set
$\sigma_{33}\approx 0$. Eq.
\ref{equation-evolution}-\ref{equation-evolution2} therefore reduce to the following system :
\ben
\label{equation2niveaux}
    \dot{\sigma}_{22}^{(0)}&=&r \sigma_{11}^{(0)}-\Gamma \sigma_{22}^{(0)} \\
    \dot{\sigma}_{11}^{(0)}&=&-r \sigma_{11}^{(0)} \
\een
whose solutions are, for $t\leq \delta T$:
\ben
\label{solution<dT}
   \sigma_{11}^{(0)}(t,t_0;1) &=& \exp[-r(t-t_0)] \\
    \sigma_{22}^{(0)}(t,t_0;1) &=& =\frac{r}{r-\Gamma} 
\left( \exp[-\Gamma(t-t_0)]-\exp[-r(t-t_0)]\right) \
\een
and for $t\geq\delta T$:
\ben
\label{solution>dT}
   \sigma_{11}^{(0)}(t,t_0;1) &=& \sigma_{11}^{(0)}(\delta T,t_0;1) \\
    \sigma_{22}^{(0)}(t,t_0;1) &=& \exp[-\Gamma(t-\delta T)]\sigma_{22}^{(0)}(\delta T,t_0;1) \
\een
The probability $P_e^{(g)}$ to emit at least one photon between two pulses
(say in interval $[0,T]$) is then:
\begin{equation}\label{pe}
  P_e^{(g)}=\Gamma \int_{0}^{T}\sigma_{22}^{(0)}(t,0;1) dt=1-\exp(-r\delta
  T)-\frac{r}{r-\Gamma}\exp(-\Gamma T)[1-\exp((\Gamma-r)\delta T)]
\end{equation}
This probability is of course increasing with the period T, which
has to be large compared to $\Gamma^{-1}$ to assure the emission
of the photon (for instance, $\exp(-\Gamma T)=5 \; 10^{-5}$ for a
10 MHz pulse train and $\Gamma^{-1} \approx 10$ns). We can therefore
set:
\begin{equation}\label{pe_approx}
 P_e^{(g)} \approx 1-\exp(-r\delta T)
\end{equation}
The probability $P_{n}^{(g)}$ to emit exactly n photons is given
by:
\begin{equation}\label{Pn(g)}
P_{n}^{(g)}=\sum_{a}\sigma_{aa}^{(n)}(T,0;1) =\int_{0}^{T}dt
\left\{1-\Gamma
\int_{t}^{T}\sigma_{22}^{(0)}(t^{'},t;1)dt^{'}\right\}\Gamma
\sigma_{22}^{(n-1)}(t,0;1)
\end{equation}
where the second equality corresponds to the
probability to emit the photon $n$ at time $t$, and no
photons within $[t,T]$. In the limit
$\exp(-\Gamma T)\rightarrow 0$, the probability 
$P_{1}^{(g)}$ is given by the following
expression, which is well-behaved when $r=\Gamma$ :
\begin{equation}\label{P1(g)}
  P_{1}^{(g)}=(\frac{r}{r-\Gamma})^{2}[\exp(-\Gamma \delta T)-\exp(-r\delta
  T)]-\frac{\Gamma r \delta T }{r-\Gamma} \exp(-r\delta T)
\end{equation}

\subsection{Non-perfect collection efficiency}
\label{sec_b}

In practice, the dipole cannot be
separated from the collection system, and the statistics of interest
is the statistics of
the detected events, rather that the one of the emission events. 
Assuming again that the initial state is the ground
state, and denoting as $\eta$ is the collection
efficiency ($\bar{\eta}=1-\eta$), the probability to collect no
photon between $[0,T]$ is :
\begin{equation}\label{no_photon}
  \Pi_{0}^{(g)}=\sum_{n=0}^{\infty}\bar{\eta}^{n}P_{n}^{(g)}
\end{equation}
Let us introduce the 
probability $\tilde{\sigma}_{aa}$ 
to reach state $|a\rangle$ without the collection of
any photon, which is given by :
\begin{equation}\label{sigma_tild}
  \tilde{\sigma}_{aa}=\sum_{n=0}^{\infty}\bar{\eta}^{n}\sigma^{(n)}_{aa}
\end{equation}
From the above definitions, we have $\Pi_{0}^{(g)}=\sum_{a}\tilde{\sigma}_{aa}$. 
Using a calculation very
similar to the beginning of this section (see eq. \ref{coef} and Appendix 2),
the linear differential system for ${\tilde{\sigma}}_{aa}$ can be shown to be :
\ben
\label{equation-evol_sigtild}
    \dot{\tilde{\sigma}}_{22} &=& =r \tilde{\sigma}_{11}-\Gamma \tilde{\sigma}_{22}
                                                +r_{d} \tilde{\sigma}_{33} \\
    \dot{\tilde{\sigma}}_{33} &=& -\Gamma_{T} \tilde{\sigma}_{33}+\beta\Gamma \tilde{\sigma}_{22}
                                                -r_{d} \tilde{\sigma}_{33}\\
    \dot{\tilde{\sigma}}_{11} &=& -r \tilde{\sigma}_{11}+\Gamma_{T}\tilde{\sigma}_{33}
                                                +\bar{\eta}\Gamma \tilde{\sigma}_{22} \
\een
The correction introduced here, compared to equations
\ref{equation-evolution}-\ref{equation-evolution2}, consists in the addition of a term 
filling the ground state with
a rate corresponding to the probability density
$\bar{\eta}\Gamma$ to emit one photon but not to collect it. This
term ensures the collection of one and only one photon. If the
initial state is the ground state, and within the approximations
of subsection \ref{sec_a} ($\beta \ll 1$, $\tilde{\sigma}_{33}=0$), this system
can be rewritten:
\ben
\label{equation_sigmatild}
    \dot{\tilde{\sigma}}_{22} &=& r \tilde{\sigma}_{11}-\Gamma \tilde{\sigma}_{22} \\
    \dot{\tilde{\sigma}}_{11} &=& -r \tilde{\sigma}_{11}+\bar{\eta}\Gamma \tilde{\sigma}_{22} \
\een
This system can easily be solved between $[0,T]$, and we find for
$\Pi_{0}^{(g)}$, in the limit $\exp(-\Gamma T)\rightarrow 0$:
\begin{equation}\label{pi_ne}
\Pi_{0}^{(g)}=\tilde{\sigma}_{11}(\bar{\eta};T,0;1)=
\bar{\eta}\tilde{\sigma}_{22}(\bar{\eta};\delta
T,0;1)+\tilde{\sigma}_{11}(\bar{\eta};\delta T,0;1)
\end{equation}
with
\ben
\label{equ_sigma_tild}
    \tilde{\sigma}_{11}(\bar{\eta};\delta T,0;1) &=&  \frac{r-\Gamma'}{r'-\Gamma'}\exp(-r'\delta T)
                        +\frac{r'-r}{r'-\Gamma'}\exp(-\Gamma'\delta T) \\
    \tilde{\sigma}_{22}(\bar{\eta};\delta T,0;1) &=& \frac{r}{r'-\Gamma'}[\exp(-\Gamma'\delta
T)-\exp(-r'\delta T)]\
\een
and

\ben
\label{prime}
r'&=&\frac{1}{2}(\Gamma+r+\sqrt{(r-\Gamma)^{2}+4\bar{\eta}r\Gamma}) \\ \nonumber
\Gamma'&=&\frac{1}{2}(\Gamma+r-\sqrt{(r-\Gamma)^{2}+4\bar{\eta}r\Gamma}) 
\een

The probability $\Pi_{0}^{(g)}$ allows us  to determine the
probability $\Pi_{e}^{(g)}=1-\Pi_{0}^{(g)}$ to collect at least
one photon. It  permits also to obtain
the probability to collect one and only one photon :
\begin{equation}\label{def_Pu}
\Pi_{1}^{(g)}=\sum_{n=1}^{\infty}n\eta\bar{\eta}^{n-1}P_{n}^{(g)}=\eta\partial_{\bar{\eta}}\Pi_{0}^{(g)}
\end{equation}
We find thus :
\ben
\label{Pu}
\Pi_{1}^{(g)}&=&\frac{\eta  r}{r'-\Gamma'} 
(1+\frac{\Gamma (2\eta r - r - \Gamma}{(r'-\Gamma')^{2}})
\left( \exp(-\Gamma' \delta T)-\exp(-r'\delta T) \right) \nonumber \\
&+& \frac{\eta  r \Gamma \delta T}{r'-\Gamma'} 
\left(\frac{r'-\eta r}{r'-\Gamma'}\exp(-\Gamma' \delta T)+\frac{\Gamma'-\eta
r}{r'-\Gamma'}\exp(-r' \delta T) \right) 
\een

These results correspond of course to the results of subsection
\ref{sec_a} if $\eta=1$.  A simpler expression can be obtained
by considering in first approximation that no more than two
photons can be emitted during the light excitation pulse. We then
have :
\begin{equation}\label{approxP}
  P_{0}^{(g)}+P_{1}^{(g)}+P_{2}^{(g)}=
1-P_{e}^{(g)}+P_{1}^{(g)}+P_{2}^{(g)}\approx 1
\end{equation}
Equation \ref{no_photon} can then be written as
\begin{equation}\label{no_photon_approx}
  \Pi_{0}^{(g)}=1-P_{e}^{(g)}+\bar{\eta}P_{1}^{(g)}+\bar{\eta}^{2}(P_{e}^{(g)}-P_{1}^{(g)})
\end{equation}
and  equation \ref{def_Pu} as
\begin{equation}\label{Pu_approx}
  \Pi_{1}^{(g)}
= \eta (P_{1}^{(g)}+2 \bar{\eta}(P_{e}^{(g)}-P_{1}^{(g)}))
\end{equation}

\subsection{Influence of the metastable state}
\label{sec_c}

In order to study the effect of the metastable state, the 3-level 
equations given above can be solved analytically in the general case,
giving lengthy and not very illuminating expressions. In physical terms,
a short intense pulse will excite the dipole just as previously, but now the 
dipole may end up in the metastable state. Thus the emission of the
single photon will be delayed by an amount depending on the time
spent in the metastable state.

For definitiveness, we shall consider here the situation where the
 transition rate $\beta \Gamma$  to the metastable level
$|3\rangle$ is weak but not completly negligible; this applies in particular
to single molecules (see next section). 
 The probability to populate this level when a transition occurs
form level $|2\rangle$ is $\frac{\beta}{\beta+1} \approx \beta$,
 so the metastable level is reached every
$(\beta P_{e}^{g})^{-1}$ light pulses in average. 
In a way similar to the approximations of subsection \ref{sec_a}, we can
neglect the probability to leave the metastable state and to reach
it again in the same cycle $[0,T]$. We can therefore neglect the
filling term $\beta\Gamma \tilde{\sigma}_{22}$ in equations
\ref{equation-evol_sigtild}, 
and the probability to stay in the metastable state in one cycle is
$\exp[-(\Gamma_M+r_{d})T]$, or for q cycles :
\begin{equation} \label{cycle}
  P_c=\exp[-(\Gamma_M+r_{d})qT]
\end{equation}
When the system reaches the metastable level, it therefore remains
shelved during a mean time $(\Gamma_M+r_{d})^{-1}$, that will be assumed
to be much larger than T. The probability
to reach this level is approximately $\beta P_e^{(g)}$ for each
excitation pulse, so the average
time it takes for the system to find itself
in the metastable state  is $T(\beta P_e^{(g)})^{-1}$. The number of
emitted photons is thus decreased by a factor
\begin{equation}\label{factor_M}
  M=\frac{T(\beta P_e^{(g)})^{(-1)}}{T(\beta P_e^{(g)})^{-1}+(\Gamma+r_{d})^{-1}}
  =\frac{(\Gamma+r_{d})T}{\beta P_e^{(g)}+(\Gamma+r_{d})T}
\end{equation}
Even if $\beta$ is small, the factor M
can thus induce a reduction of the photon flux.
Obviously this decrease in the number of emitted photons
has different statistical properties than the random
deletion considered in the previous section \cite{BFTO}.
One gets now alternatively periods where the source is ``on'', 
and periods where it is ``off''.
In a practical system, 
one may consider to use a ``deshelving'' laser \cite{DFTJKNW}
in order to increase $r_d$, and thus to maximize the duty cycle of the dipole.

\section{Discussion}

In this section the above model is used to demonstrate 
the potentiality of a single emitter to produce single photons
 when excited by a light pulse. This potentiality is evaluated by the
fractional information leakage $f_{il}$ defined in Eq. (\ref{fil}).
In particular the influence of the
duration and the energy of the pulse is investigated.
The parameters considered in the following corresponds to 
commercially available laser systems for typical emitters
such as terrylene in $p$-terphenyl \cite{FSZHW} or Nitrogen-Vacancy 
colored centers in diamond \cite{BBPG} with saturation intensity of the order of
$1$ MW/cm$^2$.
Note also that in all the plots discussed below the collection efficiency is taken as
$\eta=0.2$, which is a realistic value for an optimized passive collection 
system at room temperature.

In Fig. \ref{FilPoisF} the ability of the single emitter source 
to deliver truly single photons is compared to an
attenuated Poissonian source with the same number of empty pulses.
The  fractional information leakage $f_{il}$ is plotted 
as a function of the probability $P_e$ of emitting at least one photon. 
The quantity $P_e$ is varied by changing the pulse power while the pulse duration 
is kept constant. When the
pulse duration $\delta T$ is ten times shorter than the emitter's lifetime,
it appears that the occurence of pulses with two photons or more 
is reduced by one order of magnitude when 
a single emitter is used instead of an attenuated Poissonian source. Reducing further 
the pulse duration to $1\%$
of the emitter's lifetime improve the fractional information leakage
by another factor of $10$.

Fig. \ref{Pcst} shows the influence of the pulse duration $\delta T$
on the fractional information leakage $f_{il}$ for a given excitation peak power
(\ie for a given $r$). This would correspond to an experiment 
where the pulses are sliced up in a continuous wave laser
with fast optical modulators. 
Of course the shorter the pulse the better $f_{il}$, but when the pulse is too short
the probability $P_e$  decreases  also since 
the peak power is constant.
Note that $P_e$ can exceed the collection efficiency $\eta$ for  $\delta T$ large compared
to $\Gamma ^{-1}$.
In this case the fluorescence pulse emitted by the dipole contains 
much more than one photon, so that even after the $\eta=0.2$ attenuation the probability
of having  more than one photon remains larger than $\eta$.

In Fig. \ref{tcst} the fractional information leakage $f_{il}$ and
the probability $P_e$ of emitting at least one photon are 
plotted versus  the pulse power for a given pulse duration.  As expected short pulses
($\delta T = 0.01 / \Gamma$) require more power to reach a value of $P_e$ around 
$P_e = \eta = 0.2$, since  $P_e$ depends only on the 
pulse energy $r\delta T$. But short pulses offer a  better fractional information 
leakage $f_{il}$, owing to the fact that 
the shorter the pulse, the lower the probability of 
emitting a photon and being reexcited within the same pulse.

For usual molecules or colored centers, the excited state lifetime is of the order 
of $\Gamma ^{-1}= 10$ ns, and typical saturation intensity when focussed on a
sub-micron spot are of the order of $1$ mW. For these types of emitting dipoles 
laser pulses with $\delta T=0.1$ ns and  peak power of $1$ W (ie pulse energy of 
$0.1$ nJ) will already lead to good results.
It has to be recalled that the incoherent model used here is
valid only when the pulse duration remains larger than 
the coherence decay, that is in the picosecond range. This hypothesis prevents the use 
of extremely short and intense pulses, but is fully compatible
with the numbers just quoted.

\section{Conclusion}

We have evaluated the efficiency of a single photon source
based upon the pulsed excitation of an individual dipole,
in a regime where coherences 
are strongly damped and thus rate equations are relevant. This calculation
applies for instance to the excitation of a single molecule or
a single colored center at room temperature \cite{BBPG}.

With respect to a radiatively damped two-level system,
where either an exact $\pi$-pulse or a fast adiabatic passage is required
\cite{LK,KJRT,BLTO},
the requirement on the pulse intensity is much less stringent.
This type of system 
is thus promising for achieving an all solid state single photon source
operating at room temperature. 

This work is supported by the European IST/FET program
``Quantum Information Processing and Telecommunication'', 
project number 1999-10243 ``S4P", and by France Telecom - Centre National 
d'Etude des T\'el\'ecommunications
under the ``CTI" project number 99 1B 784.


\appendix

\section{Loss of information owing to pulses containing
two  photons or more in a quantum cryptographic scheme}
\label{appendix1}

We emphasize that $f_{il}$ defined by eq. \ref{fil} clearly gives
a physically meaningful evaluation of the single-photon character of the
pulse.
We note in particular that when $P_{n \geq 1} \ll 1$, the condition
$f_{il} < P_1/2$ is equivalent to the ``anticorrelation'' criterion
$\alpha <1$ that was introduced in ref. \cite{GRA}.
We give below a few examples that suggest to conjecture heuristically
that $f_{il}$ also gives a good indication of the
information leakage due to the multiphotonic character of the light pulses.
The quantitative evaluation of $f_{il}$, which is the main result
of this paper, obviously does not depend on the arguments given below.

For the sake of illustration, the information is supposed 
to be coded in the photon polarisation, but the following 
discussion remains valid for any types of information encoding.
A simple strategy for Eve to exploit photon pairs is to tap a fraction
$\bar{\eta}=1 - \eta$ of the beam, and to store the corresponding photons.
The polarisation of the stored photons is measured later on, when Alice
and Bob have disclosed the relevant basis information. Assuming that the
probability to get more than two photons per pulse is negligible, the
probability for Eve to catch the information is then $2 \eta \bar{\eta}
P_2$, which is maximum for $\eta = 0.5$ and takes a  value $P_2 / 2$.
The relative fraction of Bob's information which is known to Eve is then
$P_2 / P_{n \geq 1}$, which is just $f_{il}$. For attenuated light pulses,
one gets
$f_{il} \approx P_2 / P_1 \approx P_1/2$. The action of Eve creates no
polarisation
errors, and cannot be distinguished from a $50 \% $ random loss in the
transmission between Alice and Bob.

Another possible, more sophisticated
strategy for Eve is to use a fast
 polarization-insensitive
quantum non-demolition measurement \cite{GLP} of the number of photons in each pulse,
and to deflect every second photon. The polarization of the deflected photons
is measured later on, as said above. The fraction of useful bits is thus
$P_{n \geq 1}$ for Bob, $P_2$ for Eve, and the information leakage is again
$f_{il} = P_2 / P_{n \geq 1}$. This scheme introduces neither polarization
errors,
nor apparent loss. It can be nevertheless be detected by Bob
if he analyses the photon statistics of the light pulses that he receives.

In presence of high transmission losses between Alice and Bob, for instance
in long distance or free-space quantum cryptography, both methods can be
combined
to give even more powerful attacks \cite{HIGM,BHKLLMNPS}.
For instance, let us assume that Eve is able to catch the light pulses
before they go through the transmission line, and to distribute
them to Bob through her own lossless line. Using the QND set-up, Eve identifies
the pulses with more than one photon, keep one of them, and redistribute
to Bob the remaining photons in order to simulate the low efficiency $\eta_L$
of the original line between Alice and Bob. In that case, as soon as
$\eta_L < f_{il}$, Eve gets essentially all the information and remains
undetected. Though some countermeasures are possible, it is now clear that
attenuated light pulses and high transmission losses are a deadly
combination for quantum cryptography \cite{HIGM,BHKLLMNPS}.

As a numerical example, when using attenuated light pulses with a typical
value
$P_1 = 0.2$, a fraction $f_{il} = 0.1$, \ie at least $10 \%$ of the
information
may leak to Eve. By comparison,  the single photon source described in this
paper will give $P_1 = 0.1$ and $f_{il} = 0.002$ for experimentally
 reachable operating conditions ($\Gamma \delta T = 0.01$,
$r=1000 \Gamma$, overall efficiency $20 \%$, see text for definitions).
In the cryptographic situations discussed above,
the information leakage to Eve is thus reduced by a factor $50$ when using
the single
photon source.

\section{Derivation of the equation for $\tilde{\sigma}_{bb}$.}
\label{appendix2}

We will derive here the rate equations for the quantity
$\tilde{\sigma}_{bb}(\bar{\eta};t,t_0;a)$, which has been defined as :
\begin{equation}\label{sigma_tildd}
  \tilde{\sigma}_{bb}(\bar{\eta};t,t_0;a) =
\sum_{n=0}^{\infty}\bar{\eta}^{n}\sigma^{(n)}_{bb}(t,t_0;a)
\end{equation}
Using this definition, we have:
\begin{equation}\label{ap_der}
\frac{\partial}{\partial t}\tilde{\sigma}_{bb}(\bar{\eta};t,t_0;a)
= \frac{\partial}{\partial
t}\sigma^{(0)}_{bb}(\bar{\eta};t,t_0;a) +
\sum_{n=1}^{\infty}\bar{\eta}^{n}\frac{\partial}{\partial
t}\sigma^{(n)}_{bb}(\bar{\eta};t,t_0;a)
\end{equation}
We have, for every $n>0$, and from equation \ref{n_photon}:
\ben
\label{ap_der2}
 \frac{\partial}{\partial t}\tilde{\sigma}_{bb}(\bar{\eta};t,t_0;a) & 
 \frac{\partial}{\partial t}\sigma^{(0)}_{bb}(\bar{\eta};t,t_0;a) +
\sum_{n=1}^{\infty}\bar{\eta}^{n}[\Gamma\sigma_{22}^{(n-1)}(t,t_0;a)
\sigma^{(0)}_{bb}(t,t;1) \\
&+ \int^{t}_{t_0}\Gamma\sigma_{22}^{(n-1)}(t^{'},t_0;a)
\frac{\partial}{\partial t}\sigma^{(0)}_{bb}(t,t^{'};1)dt']
\een
As we obviously have $\sigma^{(0)}_{bb}(t,t;1)=\delta_{b1}$, and
using eq. \ref{rate1}, we can rewrite eq. \ref{ap_der2}:
\ben
\label{ap_fin}
\frac{\partial}{\partial
t}\tilde{\sigma}_{bb}(\bar{\eta};t,t_0;a) =&
\sum_{c}r_{cb}^{(0)}\sigma^{(0)}_{cc}(\bar{\eta};t,t_0;a)
+\sum_{n=1}^{\infty}\bar{\eta}^{n}[\delta_{b1}\Gamma\sigma_{22}^{(n-1)}(t,t_0;a)
\\
&+ \sum_{c}r_{cb}^{(0)}\int^{t}_{t_0}\Gamma\sigma_{22}^{(n-1)}(t^{'},t_0;a)
\sigma_{cc}^{(0)}(t,t^{'};1)dt^{'}]
\een
Using again eqs. \ref{n_photon} and \ref{sigma_tild}, eq. \ref{ap_fin}
becomes
\begin{equation}\label{ap}
  \frac{\partial}{\partial
t}\tilde{\sigma}_{bb}(\bar{\eta};t,t_0;a)=\delta_{b1}
\bar{\eta}\Gamma\tilde{\sigma}_{22}(t,t_0;a)+
\sum_{c}r_{cb}^{(0)} \tilde{\sigma}_{cc}(\bar{\eta};t,t_0;a)
\end{equation}
which is equivalent to eq. \ref{equation-evol_sigtild}.
Equation \ref{coef} can then be easily obtained
by setting $\bar{\eta}=1$, since
$\tilde{\sigma}_{bb}(1;t,t_0;a) = \sigma_{bb}(t,t_0;a)$.



\begin{figure}
\caption{Level scheme. The fluorescence is collected between level 
$\mid\! 2 \rangle$ and $\mid\! 1 \rangle$. Level $\mid\! 3 \rangle$
is a metastable state.}
\label{levels}
\end{figure}

\begin{figure}
\caption{Fractional information leakage $f_{il}$ versus the probability $P_e$
of emitting at least one photon for $\eta = 0.2$. 
The dashed and dotted lines correspond 
to the fluorescence of a single emitter, as described in the text, for 
different excitation pulse durations
$\delta T= 0.01 \Gamma ^{-1}$ (dashed line) and
 $\delta T= 0.01 \Gamma ^{-1}$ (dotted line). 
The thin solid line is the fractional information leakage $f_{il}$ for
a Poissonian source.}
\label{FilPoisF}
\end{figure}

\begin{figure}
\caption{Influence of the pulse duration with $\eta = 0.2$.
The pumping rate  is kept constant, $r=100 \Gamma$.
The thick line is the fractional information leakage $f_{il}$. 
The thin line is the probability $P_e$ of emitting at least one photon. 
Both quantities are plotted 
 versus the normalized duration of the exciting light pulse $\Gamma \delta T$.}
\label{Pcst}
\end{figure}

\begin{figure}
\caption{Influence of the pulse power.
All traces are plotted versus the normalized pumping rate $r/ \Gamma$
for a given pulse duration $\delta T$ with  $\eta = 0.2$.
The solid lines correspond
to the fractionnal information leakage $f_{il}$, and the dashed lines to
the probability $P_e$ of emitting at least one photon. These values are given 
for $\delta T= 0.01 \Gamma ^{-1}$ (thick  lines)
and for $\delta T= 0.1 \Gamma ^{-1}$ (thin lines).}
\label{tcst}
\end{figure}

\end{document}